\documentclass[aps,prm,twocolumn,twoside,superscriptaddress]{revtex4}
\usepackage{graphicx}
\usepackage{amsmath}
\usepackage{float}
\usepackage{amssymb}
\usepackage{placeins}
\usepackage{array}
\usepackage{subfig}
\usepackage{color}
\newcommand{\comment}[1]{}
\usepackage{pdfpages}

\usepackage[symbol]{footmisc}

\def\cbns{C$_\text{B}$}
\def\cb{C$_\text{B}$ }
\def\vncbns{V$_\text{N}$C$_\text{B}$}
\def\vncb{V$_\text{N}$C$_\text{B}$ }

\def\cn{C$_\text{N}$ }
\def\gwns{$G_0W_0$}
\def\gw{$G_0W_0$ }
\def\gns{$\Gamma$}
\def\g{$\Gamma$ }
\def\bR{\mathbf{R}}
\def\veps{\varepsilon}
\def\pans{PBE0($\alpha$)}
\def\p0a{PBE0($\alpha$)\text{}}

\begin{document}
  \setcounter{page}{1}
  \date{November 30, 2018}
  
\title{Fundamental Principles for Calculating Charged Defect Ionization Energies in Ultrathin Two-Dimensional Materials}
\author{Tyler J. Smart\footnote[2]{TJS and FW contributed equally to this work.}}
\affiliation{Department of Physics, University of California Santa Cruz, Santa Cruz, CA, 95064, USA}
\author{Feng Wu\footnotemark[2]}
\affiliation{Department of Chemistry and Biochemistry, University of California Santa Cruz, Santa Cruz, CA, 95064, USA}
\author{Marco Govoni}
\affiliation{Institute for Molecular Engineering and Materials Science Division, Argonne National Laboratory, Lemont, IL, 60439, USA}
\affiliation{Institute for Molecular Engineering, University of Chicago, Chicago, IL, 60637, USA}
\author{Yuan Ping\footnote{yuanping@ucsc.edu}}
\affiliation{Department of Chemistry and Biochemistry, University of California Santa Cruz, Santa Cruz, CA, 95064, USA}


\begin{abstract}
Defects in 2D materials are becoming prominent candidates for quantum emitters and scalable optoelectronic applications. However, several physical properties that characterize their behavior, such as charged defect ionization energies, are difficult to simulate with conventional first-principles methods, mainly because of the weak and anisotropic dielectric screening caused by the reduced dimensionality.
We establish fundamental principles for accurate and efficient calculations of charged defect ionization energies and electronic structure in ultrathin 2D materials. 
We propose to use the vacuum level as the reference for defect charge transition levels (CTLs) because it gives robust results insensitive to the level of theory, unlike commonly used band edge positions. Furthermore, we determine the fraction of Fock exchange in hybrid functionals for accurate band gaps and band edge positions of 2D materials by enforcing the generalized Koopmans' condition of localized defect states. We found the obtained fractions of Fock exchange vary significantly from 0.2 for bulk \textit{h}-BN to 0.4 for monolayer \textit{h}-BN, whose band gaps are also in good agreement with experimental results and calculated GW results. The combination of these methods allows for reliable and efficient prediction of defect ionization energies (difference between CTLs and band edge positions).
We motivate and generalize these findings with several examples including  different defects in monolayer to few-layer hexagonal boron nitride (\textit{h}-BN), monolayer MoS$_2$ and graphane. Finally, we show that increasing the number of layers of \textit{h}-BN systematically lowers defect 
ionization energies, mainly through CTLs shifting towards vacuum, with conduction band minima kept almost unchanged. 
\end{abstract}

\maketitle


\section{Introduction}
Two-dimensional (2D) materials provide the unique opportunity to scale future electronics smaller than ever believed physically possible, implying engineering 2D materials is a promising strategy that can meet the demands of future nanotechnologies \cite{Butler2013}. 
As defects play a crucial role in the optical and electronic properties of these systems, engineering of defects in 2D materials has sparked continuous interest \cite{wang,hong2017atomic,Lin2016,Dreyer2018,weston2018native,Tawfik2017}. For example, defects in \textit{h}-BN have been found to be the source of stable polarized and ultra-bright single-photon emissions at room temperature\cite{bourrellier2016bright,Abdi2018,Aharonovich170,Tran2015}. Hence, the development of our understanding of defects in 2D materials will open up further possibilities for emerging applications in quantum information and nanotechnology with much better scalability than traditional defects in 3D materials.

Unlike in their 3D counterparts \cite{Freysoldt2009,Freysoldt2014,Vinichenko2017,Kumagai2014,Komsa2013}, first-principles techniques for calculating defect properties in 2D materials still face significant challenges.  Specifically, eliminating the periodic charge interactions for charged defects in 2D materials requires a charge correction scheme that accounts for the weak and anisotropic dielectric screening of 2D systems \cite{Komsa2014,Wang2015}. Furthermore, several exchange-correlation functionals that provide accurate electronic structures for 3D bulk systems are no longer applicable to ultrathin 2D systems.  
For example, the fraction of Fock exchange ($\alpha$) in hybrid functionals can be approximated as the inverse of dielectric constant ($\varepsilon_{\infty}$) of the material \cite{alkauskas2011defect,Jonathan2014}; however, $\varepsilon_{\infty}$ cannot be uniquely defined for low dimensional systems as discussed in previous studies.\cite{Delerue2003,Hamel2008,brawand2016generalization}
Therefore, the determination of $\alpha$ in hybrid functionals for 2D materials remains an open question. On the other hand, many body perturbation theory techniques ($\textit{e.g.}$ GW approximation) give accurate quasiparticle energies such as band gaps and band positions; however, high computational cost and slow convergence with respect to empty states make the screening of many defects in 2D materials impractical with conventional implementations \cite{Qiu2016,Rasmussen2016,Thygesen2017,Huser2013,Attaccalite2011,Jornada2017}.

In our previous work \cite{feng,Sundararaman2017}, we developed an efficient and accurate method that can give reliable charge corrections for total energies and electronic states of charged defects in 2D materials \textit{without any supercell extrapolations}, and then provided accurate defect CTLs with the DFT+GW scheme \cite{Malashevich2014,Chen2013,Chen2015,Chen2017}.
Such  implementation is built on top of the WEST-code \cite{west}, Quantum-Espresso \cite{Giannozzi2017} and JDFTx \cite{Sundararaman2017b} packages. In our GW calculations, we avoided explicit inclusion of empty states and inversion of dielectric matrices \cite{west, ping2013,anh2013}, while also speeding up vacuum size convergence with a 2D Coulomb truncation \cite{Ismail-Beigi2006}. 

In this paper we will further investigate the possibilities of even less expensive but similarly reliable computational methods for both defect charge transition levels and ionization energies in 2D materials. The rest of the paper is organized as follows. In section II, we discuss the methodology, including details of computational methods in II.A. and how to compute thermodynamic charge transition levels in general in II.B. Then in section III we discuss our results where we have addressed two important issues for 2D materials: in section III.A., we determine which level of theory and which electron chemical potential reference should be used to calculate a CTL in 2D systems; in section III.B., we show how to define the fraction of Fock exchange in hybrid functionals for accurate band edges and band gaps; in the end, section III.C., we combine these two findings to obtain accurate defect ionization energies for 2D materials.


\section{Methodology}

\subsection{Computational Methods}
In this work, all structural relaxations and total energy calculations were performed using open source plane wave code Quantum-ESPRESSO \cite{Giannozzi2009} with ONCV norm-conserving pseudopotentials \cite{Hamann2013,Schlipf2015}, a wavefunction cutoff of 70 Ry and a $k$-point mesh equivalent to $12\times12\times1$ or higher in the primitive cell. 
 The GW calculations were performed using the WEST code \cite{Govoni2015}, which avoids explicit empty states and inversion of dielectric matrices. We employed the contour deformation technique for frequency integration of the self energy. The final values of GW corrections were extrapolated between $9\times 9$ and $12\times 12$ $k$-point meshes to infinite $k$-points similar to Ref.~\citenum{feng}. A two-dimensional Coulomb truncation~\cite{Ismail-Beigi2006,feng} has been applied to speed up the vacuum size convergences. The charge corrections for the total energies and eigenvalues of charged defects employed the techniques developed in Ref. \citenum{Sundararaman2017,feng}, which were implemented in the JDFTx code \cite{Sundararaman2017b,Ismail-Beigi2000,Arias1992}. More computational details can be found in supplementary materials.

\subsection{Thermodynamic Charge Transition Levels}

\begin{figure}[t]
\includegraphics[keepaspectratio=true,width=\linewidth]{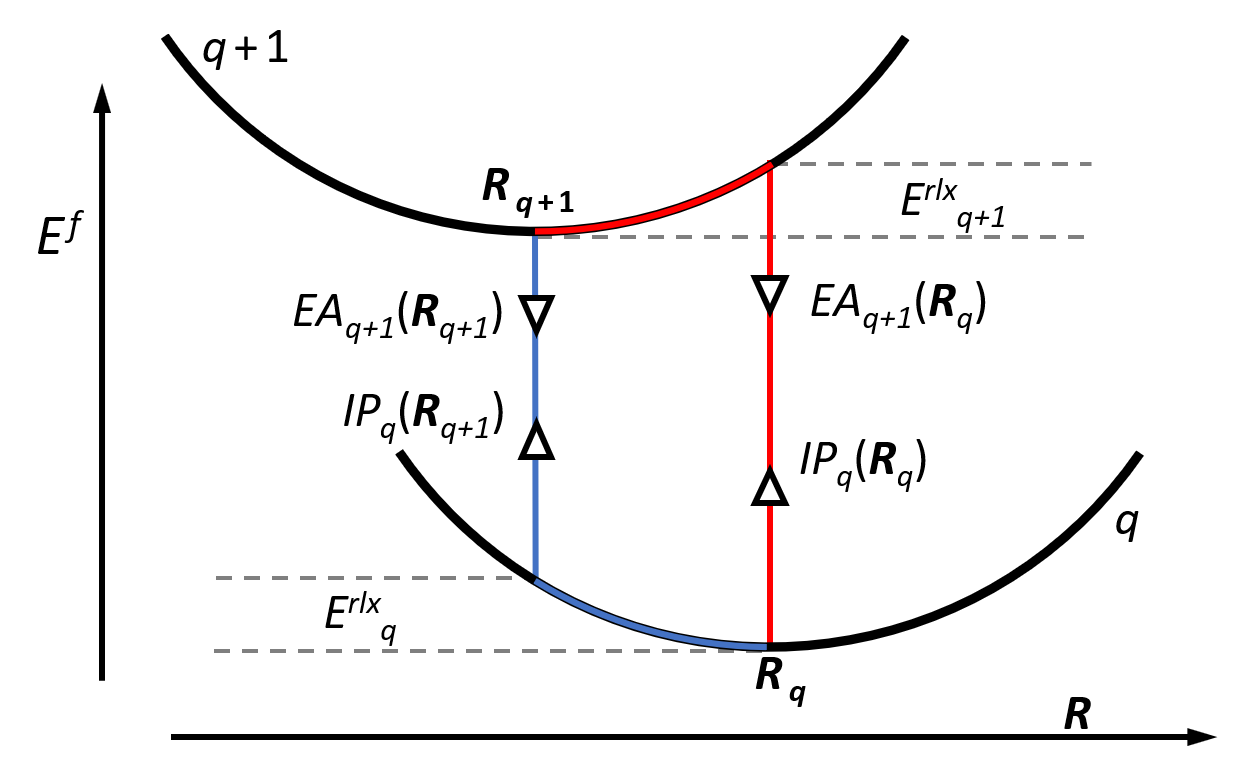}
\caption{Schematic plot of the two paths (distinguished with blue/red color) that transition from charge state $q$ to $q+1$. For each path, there is a corresponding vertical excitation, which can be computed either with EA$_{q+1}$ or IP$_q$ (noted with up/down arrowheads), as discussed in the main text.}  \label{fig:path}
\end{figure}

A thermodynamic CTL is the value of electron chemical potential $\veps_F$ at which the stable charge state of the system changes, \textit{e.g.} from $q$ to $q+1$. Therefore, CTLs are calculated through the equivalency of the formation energies $q$ and $q+1$, given by Eq. (\ref{eq:ctl0}) \cite{Freysoldt2014}.
\begin{align}
\veps_{q+1|q}&=E_q^f(\bR_q)-E_{q+1}^f(\bR_{q+1}) \nonumber\\
&=E_q(\bR_q)-E_{q+1}(\bR_{q+1})-\veps_F
\label{eq:ctl0}
\end{align}
Here $E_q^f(\bR)$ is the defect formation energy with charge $q$ and geometry $\bR$, and $\bR_q$ is the relaxed geometry of the system with charge $q$. $E_q(\bR)$ is the total energy that relates to $E_q^f(\bR)$ and $\veps_F$ following the definition of Eq. (1) in Ref. \citenum{feng}.  Diagrammatically, Eq. (\ref{eq:ctl0}) is the energy difference between two potential  surface minima in position space $\bR$, as shown in Fig. \ref{fig:path}.  

\section{Results and Discussions}
\subsection{Implementing Quasiparticle Corrections in Defect Charge Transition Levels}

In extended systems, local and semi-local functionals fail to yield accurate total energy differences between two charge states, i.e. where an electron removing (IP)/adding process (EA) is involved. An alternative approach \cite{feng} is to separate Eq.~(\ref{eq:ctl0}) into two parts: the vertical excitation energy between two charge states ($q$ and $q+1$) at the same geometry ($\bR$) (denoted as quasiparticle energies $\veps^{QP}$) and the geometry relaxation energy at a fixed charge state (denoted with $E^{rlx}$). 
In general, as ground states theory, DFT provides reliable geometry relaxation energies at a fixed charge state in many systems. For the 2D systems we tested in this work, we found the total energy difference of geometries optimized at semi-local and hybrid functionals is less than 10 meV, which leads to a negligible difference in charge transition levels (see SI Table IV). Hence, this separation allows us to accurately calculate the vertical excitation energies with a higher level of theory appropriate for non-neutral excitations, such as the GW approximation.

One can separate Eq.~(\ref{eq:ctl0}) by two possible physical pathways from $E_q^f(\bR_q)$ to $E_{q+1}^f(\bR_{q+1})$ as shown in Fig. \ref{fig:path}. One pathway (red path) occurs with a vertical excitation at $\bR_{q}$ ($E_{q+1}^f(\bR_q)-E_q^f(\bR_q)$) followed by a geometry relaxation at the charge state $q+1$ ($E_{q+1}^f(\bR_{q+1})-E_{q+1}^f(\bR_q)$), shown in Eq.~(\ref{eq:ctl1}). The other pathway (blue path) occurs through the geometry relaxation at the charge state $q$ plus a vertical excitation at $\bR_{q+1}$, corresponding to Eq.~(\ref{eq:ctl2}).
\begin{align}
\veps_{q+1|q} &= \underbrace{E_{q}^f(\bR_q)-E_{q+1}^f(\bR_q)}_{\veps^{QP}}+\underbrace{E_{q+1}^f(\bR_q)-E_{q+1}^f(\bR_{q+1})}_{E^{rlx}}  \nonumber\\
&= \veps^{QP}_{q+1|q}(\bR_q) + E^{rlx}_{q+1} 
\label{eq:ctl1}
\end{align}
\begin{align}
\veps_{q+1|q}&=\underbrace{E_q^f(\bR_q)-E_q^f(\bR_{q+1})}_{E^{rlx}}+\underbrace{E_q^f(\bR_{q+1})-E_{q+1}^f(\bR_{q+1})}_{\veps^{QP}} \nonumber\\
&= E^{rlx}_q  + \veps^{QP}_{q+1|q}(\bR_{q+1})
\label{eq:ctl2}
\end{align}
Note that all three equations (Eq. (\ref{eq:ctl0}), (\ref{eq:ctl1}), (\ref{eq:ctl2})) are theoretically equivalent. Yet, in practice they may yield sizable differences, when $\veps^{QP}$ is computed through eigenvalues at a specific level of theory instead of total energy differences.

Furthermore, the vertical excitation energies $\veps^{QP}_{q+1|q}$ in Eq.~(\ref{eq:ctl1}) and Eq.~(\ref{eq:ctl2}) can be determined from either the ionization potential of the charge state $q$ ($\text{IP}_q$) or
the electron affinity of the charge state $q+1$ ($\text{EA}_{q+1}$)
, as noted in Fig.~\ref{fig:path} with up/down arrowheads. We obtained IP and EA through eigenvalues at different levels of theory based on the Janak's theorem \cite{Janak}.
Within the framework of DFT with local and semi-local functionals, energy is a continuous and smooth functional of the number of electrons in the system $E[n]$. This results in non-linear behavior with respect to electron number and gives inconsistent eigenvalues ($\partial E / \partial n$) of the $q$ and $q+1$ systems (resulting in $\text{IP}\neq\text{EA}$) \cite{perdew1982density,cohen2008insights}. Following the discussions in Ref. \citenum{Fabien2009}, an eigenvalue between the $q$ and the $q+1$ systems can be approximated to the second order as:
\begin{equation}
  \veps^{QP}_{q+1|q}(\bR) \approx - \frac{1}{2}\left[ \frac{\partial E}{\partial n} \bigg|_{n=(q+1)^-} + \frac{\partial E}{\partial n} \bigg|_{n=q^+} \right]
  \label{eq:avg1}
\end{equation}
Therefore, to the second order in electron number, $\veps^{QP}$ is obtained by taking the average of $\text{IP}_q$ and $\text{EA}_{q+1}$:
\begin{equation}
  \veps^{QP}_{q+1|q}(\bR) = \frac{1}{2}\bigg[\text{EA}_{q+1}(\bR) + \text{IP}_q(\bR)\bigg]
  \label{eq:avg2}
\end{equation}

\begin{table}[t]
\centering
\comment{\begin{tabular}{p{1.5cm}p{1.5cm}p{1.2cm}p{1.2cm}p{1.2cm}p{1.2cm}}
\hline\hline
\multicolumn{2}{c}{Method} & \multicolumn{4}{c}{Defect} \\
 & & \cb & \vncb & C$_\text{N}$ & \vncb \\
\hline
&&&&&\\
     &    CTL             & (0/+1) & (0/+1) & (-1/0) & (-1/0) \\
   &&&&&\\   
     & Eq\ref{eq:ctl0} & -3.63  & -4.22  & -3.54  & -1.57  \\
PBE  & Eq\ref{eq:ctl1} & -3.61  & -4.29  & -3.51  & -1.66  \\
     & Eq\ref{eq:ctl2} & -3.64  & -4.33  & -3.49  & -1.67  \\
 &&&&&\\
     & Eq\ref{eq:ctl0} & -3.65  & -4.19  & -3.50  & -1.87  \\
PBE0 & Eq\ref{eq:ctl1} & -3.60  & -4.17  & -3.50  & -1.87  \\
     & Eq\ref{eq:ctl2} & -3.62  & -4.21  & -3.50  & -1.21*  \\
&&&&&\\
     & Eq\ref{eq:ctl1} & -3.40  & -4.29  & -3.74  & -1.74 \\
\gw  & Eq\ref{eq:ctl2} & -3.28  & -4.22  & -3.73  & -1.70  \\
    &&&&&\\
    \multicolumn{6}{c}{IP$_{q}(\bR_{q})$-EA$_{q+1}(\bR_{q})$} \\
     &&&&&\\
PBE && 2.68  & 2.60  & 2.75  & 2.50  \\
PBE0 && 1.15  & 1.09  & 1.13  & 1.42  \\
\gw && 0.04  & 0.20  & 0.03  & 0.19  \\
 \hline\hline
\end{tabular}}
\begin{tabular}{p{1cm}p{1.1cm}p{1.1cm}p{1.1cm}p{1.1cm}p{1.1cm}| @{\hspace{2mm}} c}
\hline\hline
\multicolumn{2}{c}{ } & \multicolumn{5}{c}{Defect} \\
\multicolumn{2}{c}{Method} & \hspace{2mm}\cb & \vncb & \hspace{2mm}C$_\text{N}$ & \vncb & V$_\text{S}$ \\
\multicolumn{2}{c}{  } &&\multicolumn{2}{c}{(\textit{h}-BN)} && (MoS$_2)$\\
\hline
&&&&&\\
     &    CTL             & (0/+1) & (0/+1) & (-1/0) & (-1/0) & (-1/0) \\
   &&&&&\\  
PBE  & Eq\ref{eq:ctl0} & -3.63  & -4.22  & -3.54  & -1.57 & -4.29  \\
     & Eq\ref{eq:ctl1} & -3.61  & -4.29  & -3.51  & -1.66 & -4.29  \\
     & Eq\ref{eq:ctl2} & -3.64  & -4.33  & -3.49  & -1.67 & -4.29  \\
 &&&&&\\
PBE0 & Eq\ref{eq:ctl0} & -3.65  & -4.19  & -3.50  & -1.87 & -4.33  \\
     & Eq\ref{eq:ctl1} & -3.60  & -4.17  & -3.50  & -1.87 & -4.32  \\
     & Eq\ref{eq:ctl2} & -3.62  & -4.21  & -3.50  & -1.21* & -4.31  \\
&&&&&\\
\gw  & Eq\ref{eq:ctl1} & -3.40  & -4.29  & -3.74  & -1.74 & -4.34 \\
     & Eq\ref{eq:ctl2} & -3.28  & -4.22  & -3.73  & -1.70 & -4.38  \\
    \multicolumn{7}{c}{ }\\
    \multicolumn{7}{c}{IP$_{q}(\bR_{q})$-EA$_{q+1}(\bR_{q})$} \\
    \multicolumn{7}{c}{ }\\
PBE && 2.68  & 2.60  & 2.75  & 2.50 & 0.947  \\
PBE0 && 1.15  & 1.09  & 1.13  & 1.42 & 0.023  \\
\gw && 0.04  & 0.20  & 0.03  & 0.19 & -0.202  \\
 \hline\hline
\end{tabular}
\caption{\label{table:path} Charge transition levels (CTLs) relative to vacuum (in eV) of multiple defects in monolayer \textit{h}-BN and V$_\text{S}$ in monolayer MoS$_2$. These values are collected via three methods (Eq. (\ref{eq:ctl0}-\ref{eq:ctl2})) at various levels of theory (PBE, PBE0, \gwns$@$PBE ). The CTLs relative to vacuum are remarkably similar. The one exception, \vncb (-1/0) at PBE0 (marked with *) incidentally has a band inversion resulting in a CTL within the valence band, breaking the reliability of Eq. (\ref{eq:ctl2}).
We also show IP$_{q}(\bR_{q})-$EA$_{q+1}(\bR_{q})$ at different levels of theory. Note that at the \gw level, this difference is $<0.2$ eV.}
\end{table}

By employing this principle in Eq.~\ref{eq:avg2}, we compared the CTL obtained with PBE, PBE0 and $G_0W_0@$PBE for three different defects in monolayer BN
and S vacancy in MoS$_2$ as shown in Table \ref{table:path}.
Here we propose to set $\veps_F$ equal to the vacuum level (determined by the electrostatic potential in the vacuum region of supercells) and use it as a reference for Eq.(\ref{eq:ctl0}).
We found this choice (opposed to the commonly used band edges) is particularly advantageous for obtaining consistent CTLs among different methods as shown in Table~\ref{table:path}.
(More computational details for $G_{0}W_{0}$ can be found in SI, with numerical techniques as in Ref. \citenum{feng}).
We note that our choice of using the vacuum level as the reference for 2D materials has similarity with the idea of using the averaged electrostatic potential as the reference level for 3D materials, as discussed in several previous studies.~\cite{alkauskas2008defect,lyons2009role,komsa2011assessing,peng2013convergence}
There are several interesting observations from Table \ref{table:path}, as follows. First, we found excellent agreement (within 0.1 eV) among Eq.~(\ref{eq:ctl0}), (\ref{eq:ctl1}) and (\ref{eq:ctl2}) for each defect at a fixed level of theory if $\veps^{QP}_{q+1|q}$ in Eq.~(\ref{eq:ctl1}) and (\ref{eq:ctl2}) is obtained through Eq.~(\ref{eq:avg2}). Second, we found the results obtained among PBE, PBE0 and $G_0W_0@$PBE are also strikingly similar (within 0.2 eV) for several defects in \textit{h}-BN as well as 
V$_\text{S}$ in MoS$_2$ which has a very different chemical bonding character from \textit{h}-BN.
This suggests that CTLs of 2D materials relative to vacuum are \textit{robust to the level of theory one chooses}. Note that the difference between $\text{IP}_q$ and $\text{EA}_{q+1}$ is more than 2 eV for PBE, reduced to $~$1 eV at PBE0 level ($\alpha=0.25$), but less than 0.2 eV at $G_0W_0$, which indicates the delocalization error present in semi-local DFT has been mostly corrected at $G_0W_0$@PBE \cite{Fabien2009}. To the best of our knowledge, there are currently no available experimental values of defect charge transition levels in ultrathin 2D materials to compare with our calculations, partially due to the difficulty of controlling and identifying the chemical composition of defects in 2D materials. However, this emphasizes the need to reliably predict defect charge transition levels in 2D materials 
from first principles, and then use them to identify the chemical composition of defects by comparing with experimental measurements, such as zero phonon lines (ZPL). We note that we compared our results with the previous theoretical studies at the corresponding level of theory referenced to the valence band maximum (VBM), and obtained overall good agreement for monolayer and bulk \textit{h}-BN as well as MoS$_2$, as shown in SI Table V. 

\begin{table*}[t]
\begin{center}
\begin{tabular}{p{1.9cm}p{2.1cm}p{2.1cm}p{2.1cm}p{2.1cm}p{2.1cm}p{2.1cm}p{2cm}}
	 \hline\hline
	 \hspace{2mm} System & \hspace{4mm} PBE & \hspace{4mm} HSE & \hspace{3mm} PBE0 & \hspace{2mm} B3PW & \hspace{1mm} \p0a & \hspace{3mm} \gw &
	 \hspace{2mm} EXP.\\
	 \hline
	\hspace{0.1mm} ML MoS$_2$
	& 1.74 $|$ K$\,\rightarrow\,$K & 2.17 $|$ K$\,\rightarrow\,$K
	& 2.85 $|$ K$\,\rightarrow\,$K & 2.58 $|$ K$\,\rightarrow\,$K
	& 2.85 $|$ K$\,\rightarrow\,$K & 2.82 $|$ K$\,\rightarrow\,$K & 2.7\cite{Krane2016MoS2,Yao2017MoS2} \\
	\hspace{0.1mm} Graphane
	& 3.57 $|$ \gns$\,\rightarrow\,$\g & 4.41 $|$ \gns$\,\rightarrow\,$\g
	& 5.06 $|$ \gns$\,\rightarrow\,$\g & 5.04 $|$ \gns$\,\rightarrow\,$\g
	& 6.54 $|$ \gns$\,\rightarrow\,$\g & 6.41 $|$ \gns$\,\rightarrow\,$\g
	& -- \\
	\hspace{0.1mm} ML BN 
	& 4.71 $|$ K$\,\rightarrow\,$K & 5.70 $|$ K$\,\rightarrow\,$\g
	& 6.33 $|$ K$\,\rightarrow\,$\g & 6.33 $|$ K$\,\rightarrow\,$\g
	& 7.34 $|$ K$\,\rightarrow\,$\g & 7.01 $|$ K$\,\rightarrow\,$\g  & -- \\ 
	\hspace{0.1mm} BL BN 
	& 4.49 $|$ T$\,\rightarrow\,$M & 5.81 $|$ T$\,\rightarrow\,$M
	& 6.46 $|$ T$\,\rightarrow\,$\g & 6.17 $|$ T$\,\rightarrow\,$M
	& 7.08 $|$ T$\,\rightarrow\,$\g & 7.00 $|$ T$\,\rightarrow\,$\g  & -- \\ 
	\hspace{0.1mm} TL BN 
	& 4.36 $|$ T$\,\rightarrow\,$M & 5.68 $|$ T$\,\rightarrow\,$M
	& 6.40 $|$ T$\,\rightarrow\,$M & 6.03 $|$ T$\,\rightarrow\,$M
	& 7.01 $|$ T$\,\rightarrow\,$\g & 6.92 $|$ T$\,\rightarrow\,$M & -- \\ 
	\hspace{0.1mm} Bulk BN
	& 4.22 $|$ T$\,\rightarrow\,$M & 5.60 $|$ T$\,\rightarrow\,$M
	& 6.28 $|$ T$\,\rightarrow\,$M & 5.91 $|$ T$\,\rightarrow\,$M
	& 6.07 $|$ T$\,\rightarrow\,$M & 6.01 $|$ T$\,\rightarrow\,$M  &6.08 $\pm$ 0.015 \\	
	 \hline
	 \hline
	\end{tabular}
      \captionof{table}{Electronic band gaps (eV) for various pristine 2D materials. In general, PBE severely underestimates the gap. Hybrid functionals HSE, B3PW, and PBE0 ($\alpha=0.25$) generally enlarge the bulk band gap, but still underestimate the gaps of ultrathin BN and graphane compared with experiments and GW approximation. Only PBE0($\alpha$) with $\alpha$ satisfying $\text{IP}_{q}=\text{EA}_{q+1}$ of localized defects (\cbns) in \textit{h}-BN yield gaps in good agreement with experiment \cite{expgap} and $G_0W_0$@PBE. Note that the spin-orbit coupling (which was not included in this calculation) will lower the band gap of MoS$_2$ by 0.1 eV~\cite{Alidoust2014}, which will bring our PBE0($\alpha$) and \gw results in even better agreement with experimental electronic band gap.}
      \label{table:gap}
\end{center}
\end{table*}

\subsection{Generalized Koopmans' Condition for the Fraction of Fock Exchange in 2D Materials}
After we obtained reliable CTLs relative to vacuum, we focused on how to calculate accurate band edge positions and band gaps of 2D materials in order to determine defect ionization energies. Using the GW approximation, we obtained an accurate quasiparticle band gap (indirect at T$\rightarrow$M) 6.01 eV for bulk h-BN (Table \ref{table:gap}), in excellent agreement with the experimental fundamental electronic gap 6.08 $\pm$ 0.015 \cite{expgap}. Meanwhile, our GW results for the band gap of bulk \textit{h}-BN (7.01 eV) and MoS$_2$ (2.82 eV), agree well with previously reported experimental values (see Table~\ref{table:gap}).
Nonetheless, GW is still computationally too demanding for defects' screening and difficult to obtain forces and optimize geometries. Therefore, the development of computationally affordable methods such as accurate non-empirical hybrid functionals for 2D materials is strongly desired. 

\begin{figure}[b]
\begin{center}
\includegraphics[keepaspectratio=true,width=0.9\linewidth]{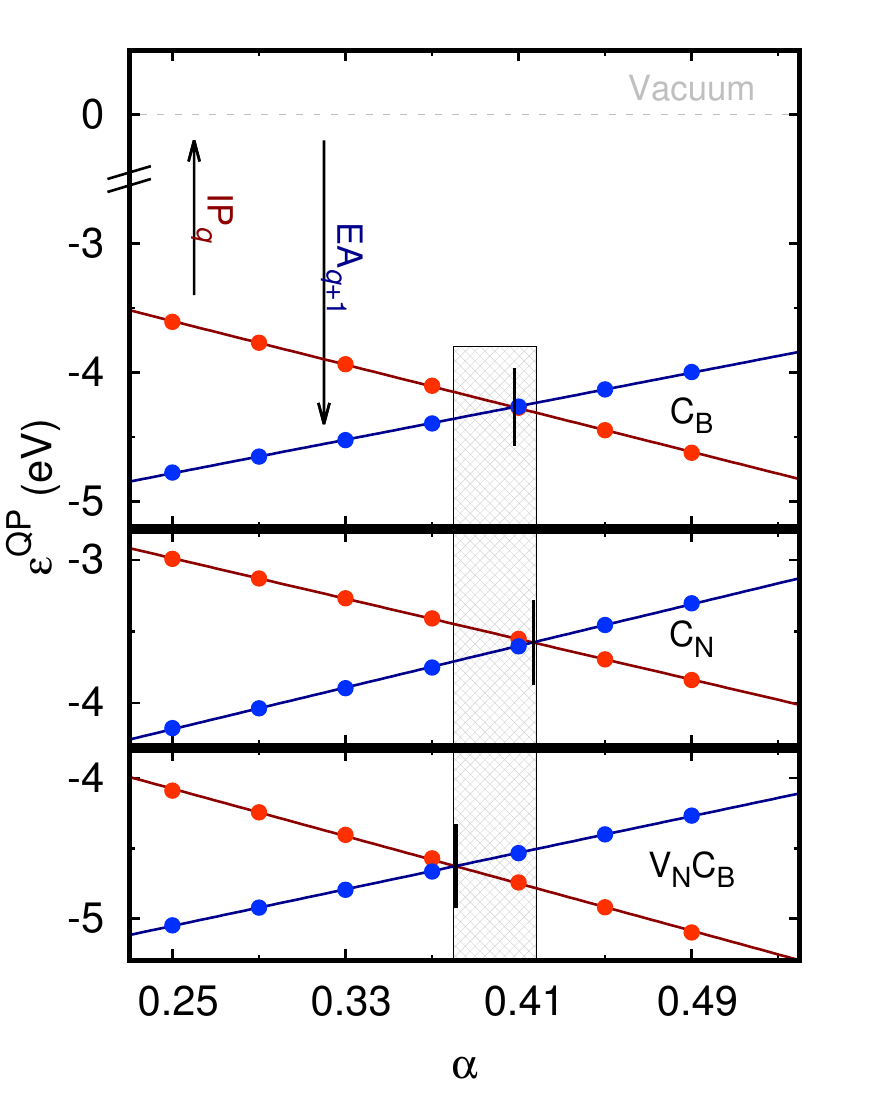}
\caption{The IP at $q=0$ and the EA at $q=+1$ for the defects \cbns , \cn and \vncb in monolayer h-BN as a function of the fraction of Fock exchange $\alpha$ for PBE0($\alpha$). The predicted exchange constant ($\alpha = 0.409$, 0.41 and 0.382, respectively) is the corresponding crossing point where $\text{EA}_{q+1}=\text{IP}_{q}$.} \label{fig:ipea}
\end{center}
\end{figure}

\begin{figure}[b]
\begin{center}
\includegraphics[keepaspectratio=true,scale=0.7]{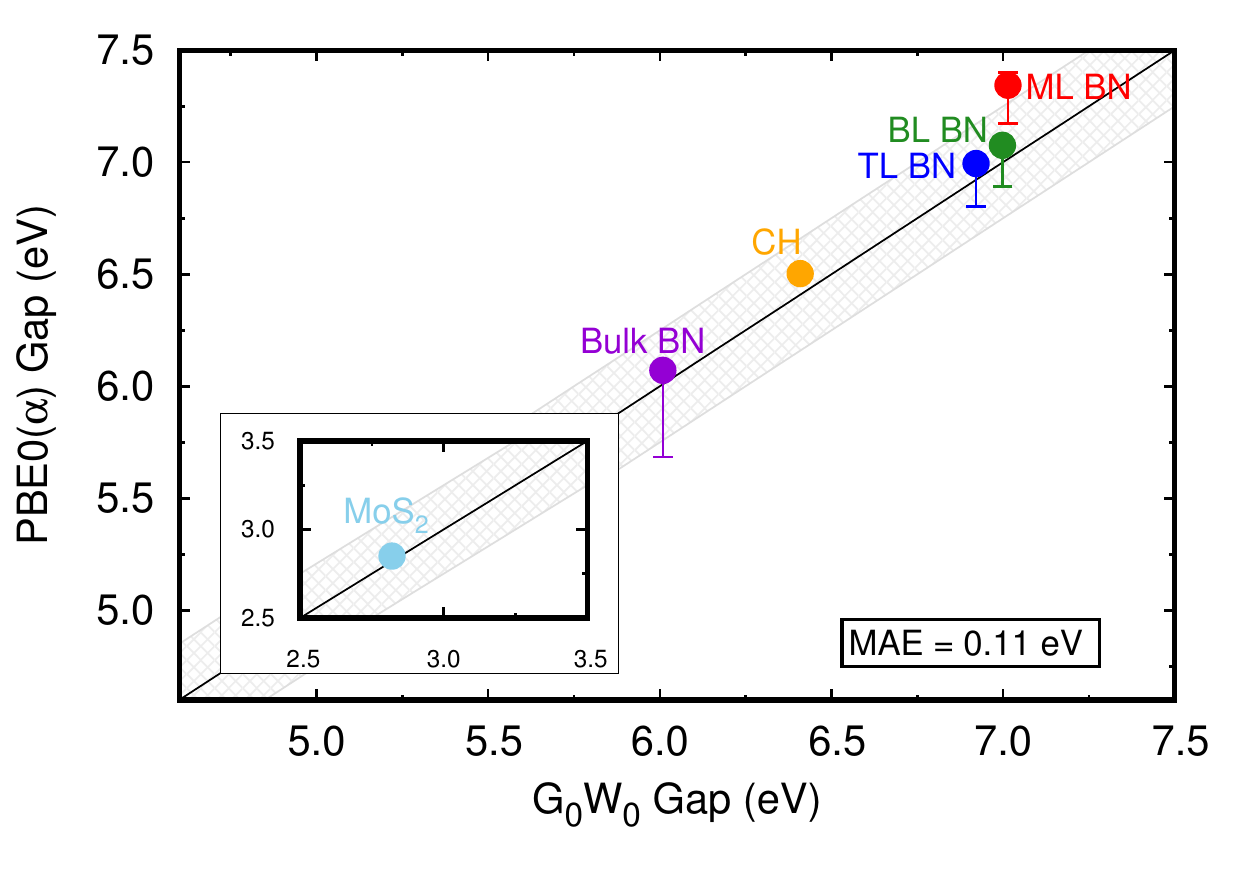}
\caption{Comparing computed band gaps of \textit{h}-BN (monolayer, bilayer, trilayer, bulk), graphane (CH), and MoS$_{2}$ with \p0a versus those computed with \gwns$@$PBE . Overall we find that our \p0a results agree very well with \gwns , yielding a MAE of 0.11 eV. The grey diagonal area highlights $\pm 0.25$ eV agreement.} \label{fig:p0a}
\end{center}
\end{figure}

\begin{table}[H]
\centering
\begin{tabular}{p{2.2cm}p{1.8cm}p{1.8cm}p{1.8cm}}
\hline\hline
 System & Defect & $\alpha$ & Gap \\
\hline
ML BN 
 & \cb   & 0.409 & 7.344 \\
 & \cn   & 0.418 & 7.401 \\
 & \vncb & 0.382 & 7.174 \\
BL BN   
 & \cb   & 0.347 & 7.075 \\
 & \cn   & 0.351 & 7.101 \\
 & \vncb & 0.318 & 6.892 \\
TL BN  
 & \cb   & 0.324 & 6.994 \\
 & \cn   & 0.326 & 7.007 \\
 & \vncb & 0.298 & 6.801 \\
Bulk BN  
 & \cb   & 0.225 & 6.071 \\
 & \cn   & 0.227 & 6.087 \\
 & \vncb & 0.178 & 5.684 \\
Graphane
 & B$_\text{C}$ & 0.467 & 6.503 \\
 & N$_\text{C}$ & 0.473 & 6.541 \\
 \hline\hline
\end{tabular}
\caption{\label{table:ipea} Predicted fraction of Fock exchange for use in the \p0a functional based on the IP$_{q}=$EA$_{q+1}$ condition. Note that for \textit{h}-BN the corresponding MAE compared with \gw results are 0.14 eV, 0.16 eV and 0.18 eV for \cbns , \cn and \vncbns , respectively.}
\end{table}

The generalized Koopmans' condition has been most commonly used to determine the appropriate fraction of Fock exchange ($\alpha$) for molecules and molecular crystals \cite{Atalla2016, Thomas2014,Abrams2012,sai2011,cohen2007,Pinheiro2015,Jie2016}. 
This principle has been successfully extended to defects in bulk semiconductors \cite{miceli} through the enforcement of this condition (\textit{i.e.} $\text{EA}_{q+1}=\text{IP}_{q}$) on defects in bulk semiconductors to obtain $\alpha$ and in turn predict accurate electronic structure of the corresponding pristine bulk systems.
The fundamental assumption is that the optimized $\alpha$ depends on the long range screening of the system and not on the nature of the probe defects. This condition is also valid for deep defects in 2D materials, where defect wavefunctions are well localized like molecular orbitals in the supercells, and their  contribution to dielectric screening is negligible compared to the crystal environment.  
Another advantage of applying this condition to 2D systems is that both $\text{EA}_{q+1}$ and $\text{IP}_{q}$ can be exactly referenced to vacuum.
In order to validate the applicability of the generalized Koopmans' condition to 2D materials, we used the defect \cb as a probe to determine $\alpha$ for \textit{h}-BN (B$_\text{C}$ (boron substitution of carbon) for graphane and V$_\text{S}$ (sulfur vacancy) for MoS$_2$). This method gives $\alpha$ of 
0.409, 0.347, 0.324, 0.225 for monolayer, bilayer, trilayer and bulk \textit{h}-BN, respectively as shown in Table~\ref{table:ipea}. Note that the $\alpha$ value 0.225 for bulk h-BN, agrees well with the predicted $\alpha$ from the inverse of high frequency dielectric constant ($\alpha = 1/\veps_{\infty} \approx 0.2$) \cite{dielectric1}, which supports the assumption that long-range screening determines $\alpha$. We also investigated other defects C$_{\text{N}}$ and {\vncbns} as probes of $\alpha$ as shown in Fig.~\ref{fig:ipea} (their corresponding electronic structure can be found in SI). 

Interestingly, we found that IP$_q$ and EA$_{q+1}$ from Kohn-Sham eigenvalues varied linearly with $\alpha$. 
Fig. \ref{fig:ipea} shows this linearity for three defects in monolayer \textit{h}-BN, and three defects predict similar $\alpha$, which justifies the insensitivity of $\alpha$ to the explicit defect. Note that defects with localized wavefunctions such as atomic substitutions (\cb) determine more accurate $\alpha$ compared with less localized defects such as {\vncbns}, because the former's contribution to dielectric screening is negligible and the screening is mostly determined by the crystal. 
It is also notable that the slopes of  IP$_q$ and EA$_{q+1}$ are opposite but nearly equal, explaining why the average of IP$_q$ and EA$_{q+1}$ as $\veps^{QP}$ for CTL in Eq. (\ref{eq:avg2}) works well (as shown in Table \ref{table:path}).

Most commonly, two-dimensional systems are synthesized with a few layers of the material, therefore understanding the effect of increasing thickness is essential to connect with realistic experiments. As such, we have computed the band gaps of monolayer, bilayer, trilayer and bulk \textit{h}-BN, as well as graphane and MoS$_2$, with several hybrid functionals including HSE, PBE0, B3PW and \p0a (with $\alpha$ predicted earlier), and \gwns @PBE as shown in Table \ref{table:gap} (the experimental photoemission gaps are also shown). As anticipated, PBE strongly underestimated monolayer \textit{h}-BN band gap: 4.71 eV with a direct transition at the K point. With any level of theory beyond PBE, monolayer \textit{h}-BN is predicted to have a larger, indirect gap from K to \gns. In accordance with quantum confinement, we observed that the band gaps of \textit{h}-BN obtained at B3PW,  \pans , and \gw show a sharp increase at ultrathin BN (monolayer to trilayer) compared to bulk BN (in agreement with a previous study \cite{wickramaratne2018monolayer}). However, HSE and PBE0 provide almost the same band gaps between ultrathin and bulk BN. This is because there is a severe change in the dielectric screening from monolayer to bulk, and a different portion of Fock exchange must be instilled. 

\begin{figure*}
\begin{center}
\begin{picture}(500,180)
\put(0,10){\subfloat[\hspace{2mm}PBE\hspace{1mm}]{\includegraphics[keepaspectratio=true,scale=0.7]{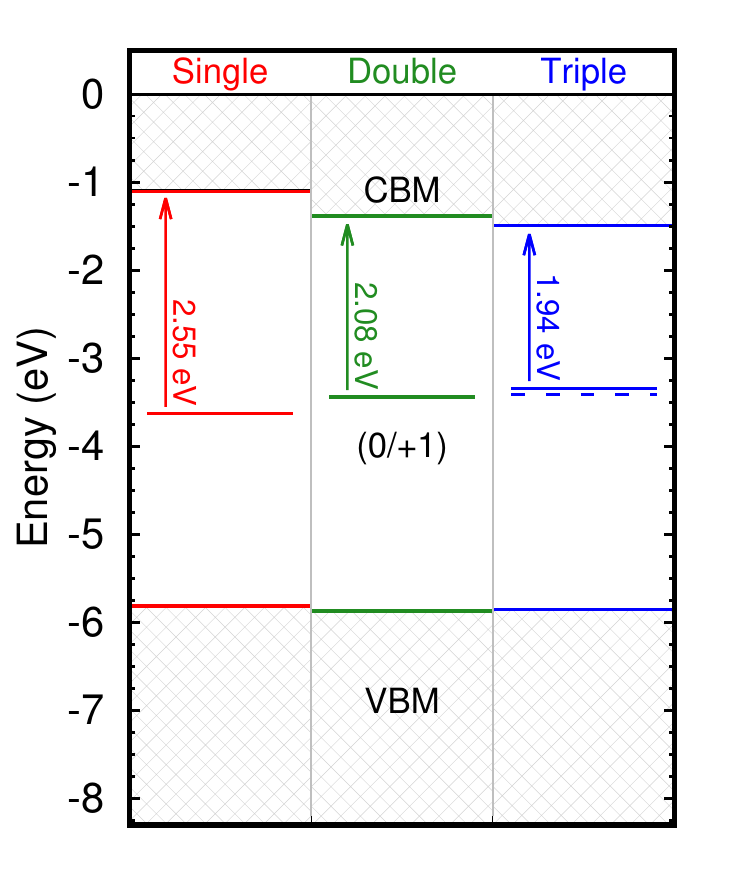}}}
\put(120,10){\subfloat[\hspace{2mm}HSE\hspace{1mm}]{\includegraphics[keepaspectratio=true,scale=0.7]{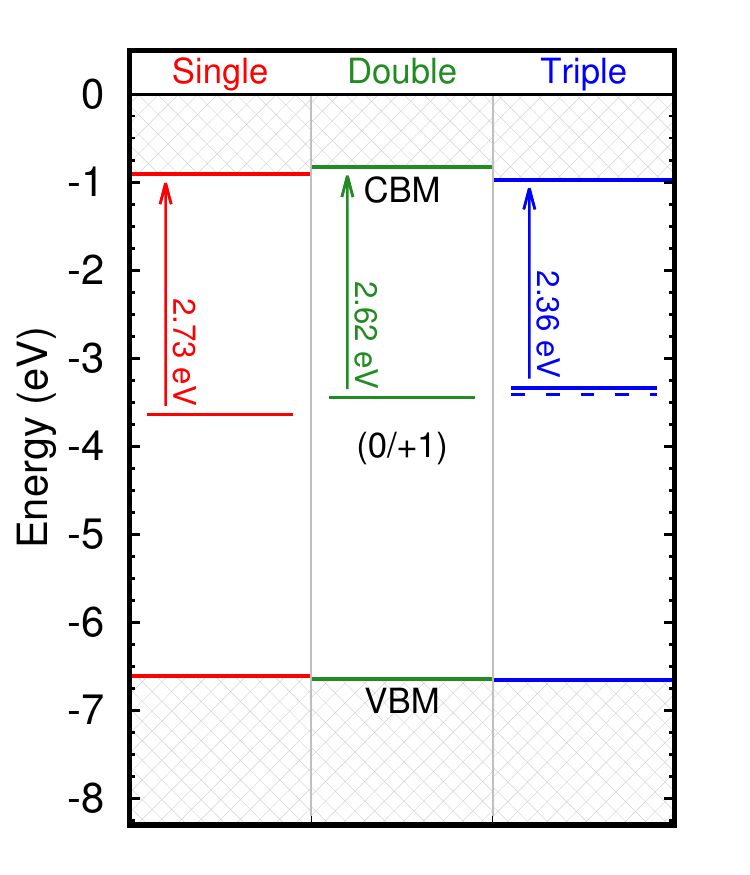}}}
\put(240,10){\subfloat[\hspace{0mm}PBE0($\alpha$)]{\includegraphics[keepaspectratio=true,scale=0.7]{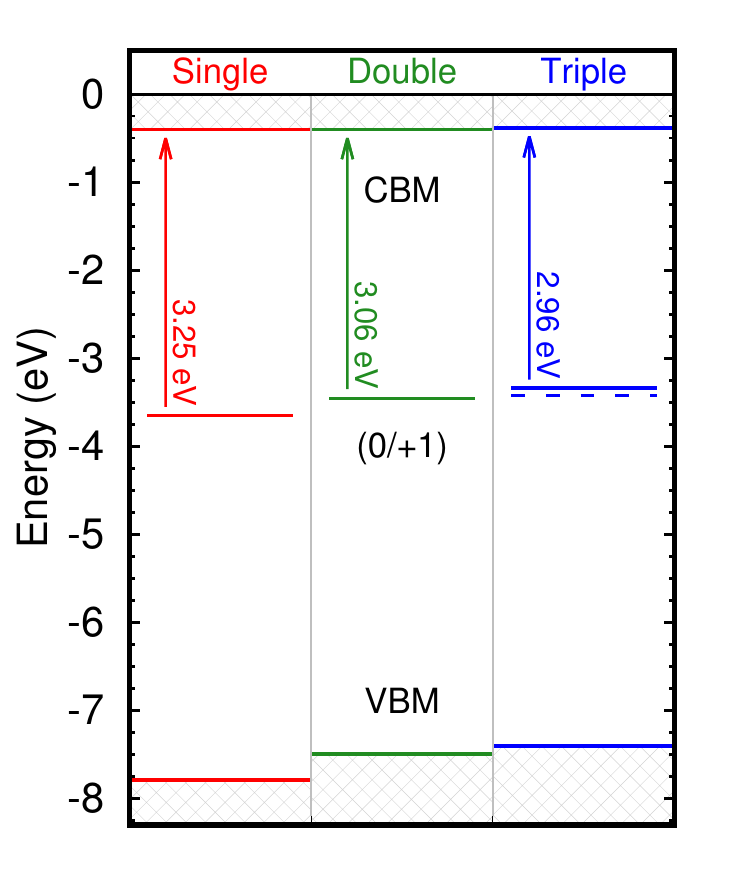}}}
\put(360,10){\subfloat[\hspace{2mm}\gw\hspace{1mm}]{\includegraphics[keepaspectratio=true,scale=0.7]{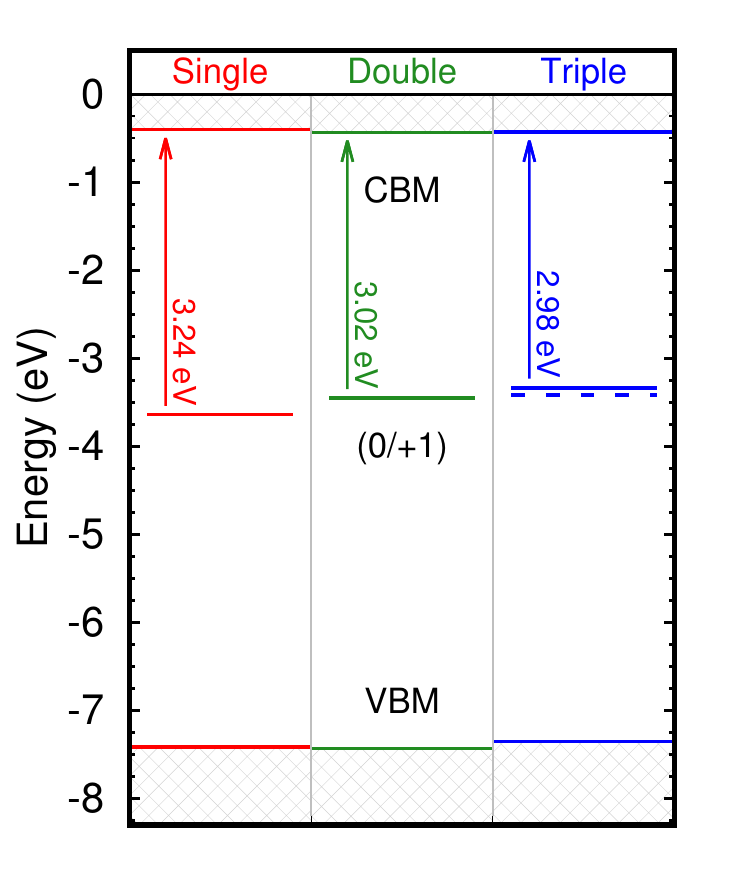}}}
\put(305,10){\includegraphics[keepaspectratio=true,scale=0.3]{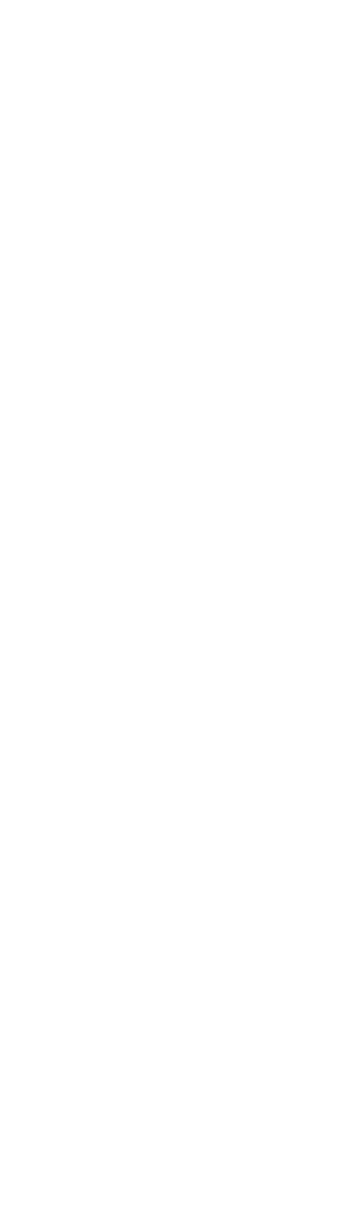}}
\put(240,10){\includegraphics[keepaspectratio=true,scale=0.7]{cbctl_pbe0ie-eps-converted-to.pdf}}
\put(185,10){\includegraphics[keepaspectratio=true,scale=0.3]{white.png}}
\put(120,10){\includegraphics[keepaspectratio=true,scale=0.7]{cbctl_hse-eps-converted-to.pdf}}
\put(65,10){\includegraphics[keepaspectratio=true,scale=0.3]{white.png}}
\put(0,10){\includegraphics[keepaspectratio=true,scale=0.7]{cbctl_pbe-eps-converted-to.pdf}}
\end{picture}
\caption{Charge transition level \cb (+1/0) in \textit{h}-BN with varying number of layers at different levels of theory. The red, green, blue lines represent the band edges and defect charge transition levels for monolayer, bilayer and trilayer  \textit{h}-BN, respectively. The dashed blue line represents the charge transition level of the defect in the middle layer of the trilayer \textit{h}-BN. Defect charge transition levels gradually become shallower with lower ionization energies while increasing the number of layers (ionization energies are written adjacent to arrows from the CTL to CBM). Note that the defect CTLs are very similar relative to vacuum between different levels of theory.}  \label{fig:ctl2}
\end{center}
\end{figure*}

Using \p0a we obtained results consistent with quantum confinement and in best agreement with our \gw calculations with a MAE of 0.11 eV (Fig.~\ref{fig:p0a}). In addition, the B3PW functional \cite{becke,bill} provided a more accurate bulk BN band gap than PBE0 and HSE but still underestimated the band gaps of ultrathin BN. 
In graphane, \p0a (with a predicted exact exchange of 0.473) leads to a gap of 6.54 eV, in agreement with \gwns @PBE, 6.41 eV. Coincidentally, MoS$_2$ has a predicted exchange of 0.250 (the default of the PBE0 scheme) and yields a gap of 2.85 eV in great agreement with \gwns @PBE, 2.82 eV. 
Therefore, the direct/indirect transitions and magnitude of the gaps from bulk to monolayer are provided accurately solely with \p0a and \gwns. 
In brief, the results shown in Table \ref{table:gap} validate our method for determining accurate fundamental band gaps for 2D materials from first-principles. We note that calculated band edge positions relative to vacuum are also similar at \p0a and \gw as shown in Fig.~\ref{fig:ctl2} and SI.

\subsection{Defect Ionization Energies in 2D Materials}
\begin{figure}[t]
\begin{center}
\includegraphics[keepaspectratio=true,scale=0.9]{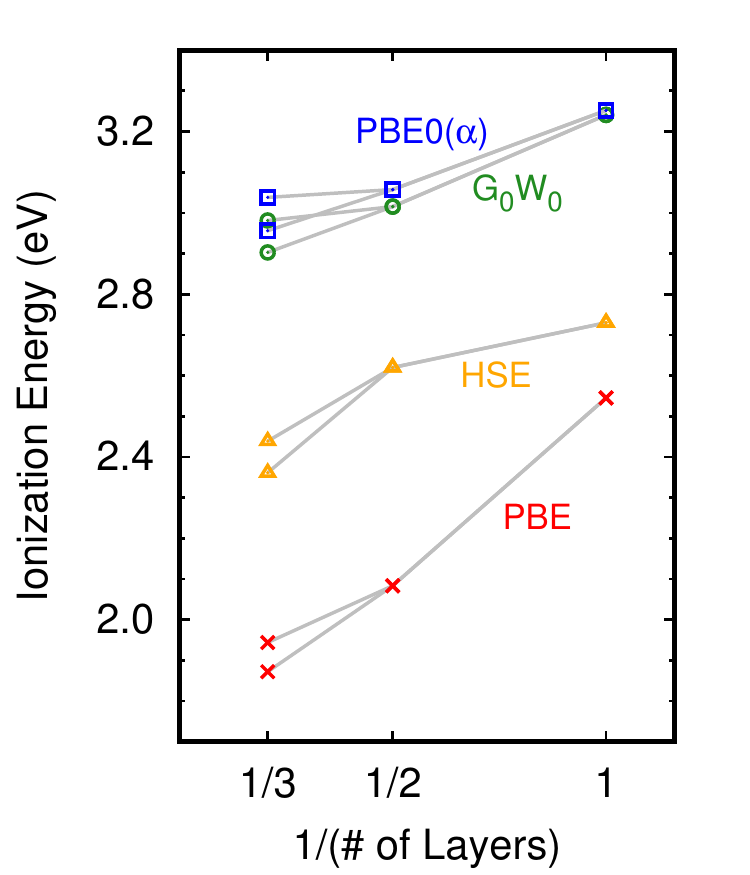}
\caption{Ionization energies of \cb in \textit{h}-BN with varying number of layers. It is observed that ionization energies decrease monotonically with increasing number of layers. Note that \p0a and \gw give results in excellent agreement.} \label{fig:ie}
\end{center}
\end{figure}

Finally, we can obtain the defect ionization energies based on the methods proposed earlier for CTLs and band edge positions relative to vacuum.
For example, CTLs and ionization energies for \cb in \textit{h}-BN computed at PBE, HSE, \p0a and $G_0W_0$ levels of theory as a function of number of layers are shown in Fig.~\ref{fig:ctl2}. Consistent with the findings in Table \ref{table:path}, CTLs changed less than 0.1 eV across different theoretical methods relative to vacuum. Interestingly, no clear trend and only small difference have been found in the band edge positions of \textit{h}-BN from monolayer to trilayer. These results illustrate that one just needs to correct the band edge positions of pristine \textit{h}-BN with \p0a or \gwns , and use CTLs determined from DFT with semi-local functionals, then the difference of the two yields accurate defect ionization energies. On another note, we found there is a clear monotonic decrease in the ionization energies of defects in \textit{h}-BN with increasing number of layers (the ionization energy is placed adjacent to the corresponding arrow in Fig.~\ref{fig:ctl2}). As shown in Fig.~\ref{fig:ie}, the ionization energy of the \cb defect in \textit{h}-BN is lowered with increasing number of layers. This effect can be understood as a result of increased dielectric screening with more layers of \textit{h}-BN, and is consistent with the effect of dielectric environments on the ionization energies of MoS$_2$ \cite{mos2}. Due to accurate band gaps and the insensitivity of charge transition levels to the level of theory as discussed earlier, the \p0a results agree well with \gw@PBE for the ionization energies of \cb in \textit{h}-BN with different number of layers(blue and green points in Fig.~\ref{fig:ie}).
	
\section{Conclusion}

In summary, we have established fundamental principles to reliably and efficiently compute ionization energies for defects in 2D materials. 
Specifically, band edge positions of the pristine systems should be computed with our proposed \p0a hybrid functional or GW approximations. We note that we obtained a fraction of Fock exchange $\alpha$ from 0.2 (bulk \textit{h}-BN) to 0.4 (monolayer \textit{h}-BN) by enforcing the generalized Koopmans' condition.
Meanwhile, we have demonstrated the insensitivity of CTL's computed by various functional choices as well as \gwns , when the CTL's are referenced to vacuum. Therefore, the defect CTL may be obtained reliably by standard DFT with semi-local functional, when it is calculated relative to vacuum.  
We have successfully applied the proposed methods for a variety of defects from monolayer to trilayer \textit{h}-BN, graphane and MoS$_2$. The combination of these methods will allow for reliable prediction and validation of defect ionization energies in two-dimensional materials, which can be potentially used to identify the chemical composition of defects in 2D materials through comparing with experimental measurements. We also demonstrated that defect ionization energies decreased with increasing number of layers in \textit{h}-BN, due to enlarged dielectric screening. Our findings in this work suggest efficient and accurate methods to compute defect ionization energies and electronic structures in 2D materials, which can be applied to screening new promising defects for quantum information and optoelectronic applications.



\section*{Acknowledgment}
We thank Giulia Galli and Chris Van de Walle for  helpful discussions. This work is supported by NSF award DMR-1760260 and DMR-1747426, and Hellman Fellowship. T.J.S. acknowledges financial support from a GAANN fellowship. M.G. is supported by MICCoM, as part of the Computational Materials Sciences Program funded by the U.S. Department of Energy, Office of Science, Basic Energy Sciences, Materials Sciences and Engineering Division. This research used resources of the Center for Functional Nanomaterials, which is a US DOE Office of Science Facility, and the Scientific Data and Computing center, a component of the Computational Science Initiative, at Brookhaven National Laboratory under Contract No. DE-SC0012704, the National Energy Research Scientific Computing Center (NERSC), a DOE Office of Science User Facility supported by the Office of Science of the US Department of Energy under Contract No. DEAC02-05CH11231, the Extreme Science and Engineering Discovery Environment (XSEDE), which is supported by National Science Foundation Grant No. ACI-1548562 \cite{xsede}, and the Argonne Leadership Computing Facility, which is a DOE Office of Science User Facility supported under Contract DE-AC02-06CH11357.

\onecolumngrid


\section*{Supplementary material}

# Supporting Information for "Fundamental Principles for Calculating Charged Defect Ionization Energies in Ultrathin Two-Dimensional Materials"

Tyler J. Smart[†],[1] Feng Wu[†],[2] Marco Govoni,[3,4] and Yuan Ping[*2]

[1]*Department of Physics, University of California Santa Cruz, Santa Cruz, CA, 95064, USA*

[2]*Department of Chemistry and Biochemistry, University of California Santa Cruz, Santa Cruz, CA, 95064, USA*

[3]*Institute for Molecular Engineering and Materials Science Division,*

*Argonne National Laboratory, Lemont, IL, 60439, USA*

[4]*Institute for Molecular Engineering, University of Chicago, Chicago, IL, 60637, USA*

(Dated: November 30, 2018)

## I. COMPUTATIONAL METHODS

In this work, all structural relaxations and total energy calculations were performed using open source plane wave code Quantum-ESPRESSO [1] with Perdew-Burke-Ernzerhof (PBE) [2] exchange-correlation functional, ONCV norm-conserving pseudopotentials [3, 4], a wavefunction cutoff of 70 Ry and a $k$-point mesh equivalent to $12 \times 12 \times 1$ or higher in the primitive cell. The vacuum between periodic images along non-periodic direction is at least 30 Bohr.

Once the structural parameters were determined, we performed a separate single-point calculation using a wavefunction cutoff of 45 Ry and hybrid functionals including HSE, B3PW91, PBE0 and PBE0($\alpha$) with a sufficient $k$-point mesh as large as $36 \times 36 \times 1$. Note that for hybrid functional calculations, the Gygi-baldereschi trick [5] plus the extrapolation at $G = 0$ term [6] has been applied for treating the Coulomb potential divergence as implemented in Quantum-ESPRESSO. The band gap is determined from the difference between valence band maximum (VBM) and conduction band minimum (CBM). If the k-point of VBM or CBM is not included in the $k$-point mesh, it is interpolated between eigenvalues of the same band of nearby $k$-points.

A single point calculation using a wavefunction cutoff of 45 Ry and PBE functional was performed as the starting point for $G_0W_0$ calculations. The GW calculations were performed using the WEST code [7]. We employed the contour deformation technique for frequency integration of the self energy. For the dielectric matrix calculation, the number of eigenpotentials ($N_{\text{PDEP}}$) was chosen to be $3N_{\text{electron}}$, and we used $4N_{\text{electron}}$ to validate its convergence. The final values of GW corrections were extrapolated between $9 \times 9$ and $12 \times 12$ $k$-point meshes to infinite $k$-points similar to Ref. 8. A two dimensional Coulomb truncation [8, 9] has been applied to the correlation part of the self energy (including the dielectric matrix) and the Gygi-baldereschi trick [5] plus the extrapolation at $G = 0$ term [6] has been applied to the exchange part of the self energy. We found this approach provides excellent vacuum size convergences and a correct convergence limit with a converged k point mesh.

The charge corrections for the total energies and eigenvalues of charged defects employed the techniques developed in Ref. 8, 10, which were implemented in the JDFTx code [11–13] (computed position dependent dielectric function profiles are shown in SI Fig. 1). Dielectric profiles are computed by applying finite electric fields following the

---

procedure discussed in Ref. 8, with a smearing width of 1 Bohr (smearing widths of 0.5 to 4.0 Bohr yield identical charge corrections).The correction to eigenvalues used the simple relation to the total energy corrections via a prefactor of $-2/q$ [14].

## II. DIELECTRIC PROFILES

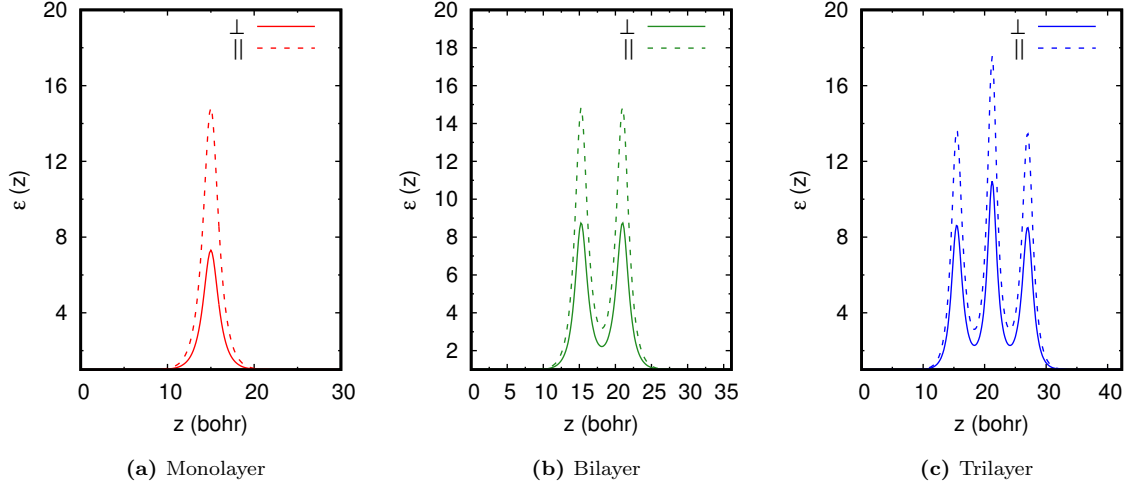

**(a)** Monolayer      **(b)** Bilayer      **(c)** Trilayer

**SI Figure 1.** Dielectric profiles of ultrathin $h$-BN. Dielectric profiles are computed by applying finite electric fields following the procedure discussed in Ref. 8, with a smearing width of 1 Bohr. "⊥" means along the perpendicular direction, and "∥" means along the parallel direction to the plane, respectively.

## III. CONVERGENCE TESTS

| Functional | Vac (Bohr) | $E_{\mathrm{vac}}$ (eV) | VBM (eV) |
|---|---|---|---|
| PBE | 30 | 2.205 | -5.822 |
| | 35 | 1.890 | -5.822 |
| | 40 | 1.654 | -5.822 |
| PBE0 | 30 | 2.206 | -6.976 |
| | 35 | 1.892 | -6.976 |
| | 40 | 1.656 | -6.975 |

SI Table I: Vacuum convergence of the valence band maximum (referenced to the vacuum level) without 2D coulomb cutoff. Here the vacuum level $E_{\mathrm{vac}}$ is computed as the average electrostatic potential in the vacuum region. The VBM is computed referenced to the vacuum level.

| Functional | Vac (bohr) | $E_{CB}$ (eV) | $E_{CB+1}$ (eV) | CCC (eV) | $\varepsilon_F$(eV) | CTL (eV) |
|---|---|---|---|---|---|---|
| PBE | 30 | -12692.2306 | -12691.3834 | 0.602 | 2.205 | -3.654 |
| | 35 | -12692.2301 | -12690.9703 | 0.507 | 1.890 | -3.657 |
| | 40 | -12692.2303 | -12690.6356 | 0.409 | 1.654 | -3.658 |
| PBE0 | 30 | -12667.3436 | -12666.5169 | 0.607 | 2.206 | -3.640 |
| | 35 | -12667.3594 | -12666.1198 | 0.511 | 1.892 | -3.642 |
| | 40 | -12667.3442 | -12665.7688 | 0.413 | 1.656 | -3.645 |

SI Table II: Vacuum convergence of charge transition levels without Coulomb cutoff. This shows that the charge transition levels (CTLs) are sufficiently converged (within 0.01 eV) at 30 Bohr, without Coulomb cutoff. Here CCC stands for charged cell correction developed in Ref. 8 and $\varepsilon_{+1/0} = E_{\mathrm{C_B}} - (E_{\mathrm{C_{B+1}}} + \mathrm{CCC}) - \varepsilon_F$, where $\varepsilon_F$ is the vacuum level $E_{vac}$ in Table I as we discussed in the main text.

| Functional | Supercell | kpoint | CTL (eV) |
|---|---|---|---|
| PBE | $6 \times 6$ | $2 \times 2$ | -3.65 |
| | $9 \times 9$ | $\Gamma$ | -3.62 |
| PBE0 | $6 \times 6$ | $2 \times 2$ | -3.64 |
| | $9 \times 9$ | $\Gamma$ | -3.62 |

SI Table III: Convergence of charge transition levels of $C_B$ defect in monolayer $h$-BN with lateral size. Small changes ($< 0.1$ eV) in the CTLs show that a $6 \times 6$ supercell size sufficiently reduced spurious in-plane interactions.

| System | Defect | $\Delta E_q^{rlx}$ (eV) | $\Delta E_{q+1}^{rlx}$ (eV) | $\mathrm{CTL_{PBE}}$ (eV) | $\mathrm{CTL_{PBE0}}$ (eV) |
|---|---|---|---|---|---|
| ML BN | $C_B$ (0/+1) | -0.0046 | -0.0009 | -3.65 | -3.65 |
| | $C_N$ (-1/0) | -0.0005 | -0.0047 | -3.50 | -3.50 |
| | $V_N C_B$ (0/+1) | -0.0143 | -0.0049 | -4.19 | -4.20 |

SI Table IV: In this work, defect calculations were relaxed with PBE and then this geometry was used for single-point calculations with hybrid functionals. This table displays the calculated differences in the value of $E^{rlx}$ for a few defects in monolayer $h$-BN if these systems are re-relaxed at the PBE0 level. The corresponding charge transision level, $\mathrm{CTL_{PBE}}$ (single point PBE0 calculation i.e. using PBE geometry), and $\mathrm{CTL_{PBE0}}$ (fully relaxed PBE0 calculation), show negligible differences.

## IV.   BAND STRUCTURES OF *h*-BN AND GRAPHANE

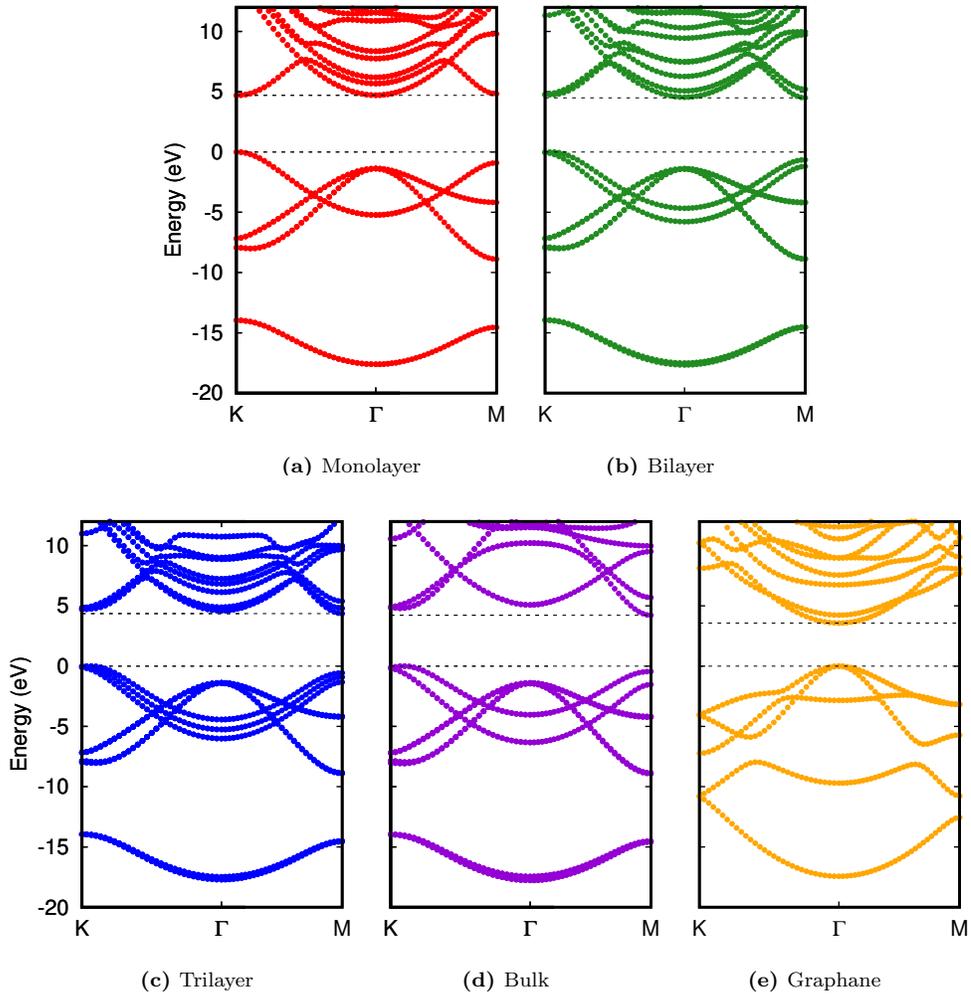

**(a)** Monolayer

**(b)** Bilayer

**(c)** Trilayer

**(d)** Bulk

**(e)** Graphane

**SI Figure 2.** PBE band structure of *h*-BN (monolayer (a), bilayer (b), trilayer (c), and bulk (d)) along with the PBE band strucutre of graphane (e).

## V. DEFECT CHARGE TRANSITION LEVELS

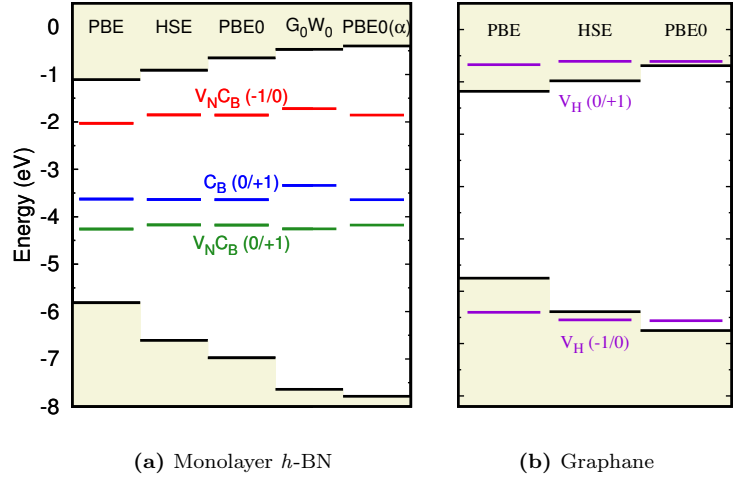

**(a)** Monolayer *h*-BN

**(b)** Graphane

**SI Figure 3.** Defect charge transition levels in monolayer *h*-BN and graphane systems. The upper and lower grey areas represent the CB and VB, respectively, computed at PBE, HSE, PBE0, $G_0W_0$, and PBE0($\alpha$).

| System | Method | Defect | This Work CTL (eV) | Other Work CTL (eV) |
|---|---|---|---|---|
| ML *h*-BN | PBE | $V_N$ (0/+1) | 1.81 | 2.08 [15] |
| | | $V_N C_B$ (0/+1) | 1.56 | 1.44 [15] |
| | HSE | $V_N C_B$ (-1/0) | 4.76 | 4.68 [16] |
| | | $V_N C_B$ (0/+1) | 2.44 | 2.57 [16] |
| Bulk *h*-BN | HSE | $C_B$ (0/+1) | 3.68 | 3.71 [17] |
| | | $C_N$ (-1/0) | 3.05 | 3.19 [17] |
| $MoS_2$ | PBE | $V_S$ (+1/0) | -0.34 | -0.3 [18] |
| | | $V_S$ (0/-1) | 1.65 | 1.7 [18] |

SI Table V: Various charge transition levels computed within this work compared with previously published values. These charge transition levels are given with respect to the valence band maximum of the pristine system in order to compare with other works.

## VI. $G_0W_0$ CONVERGENCE

| System | $N_{\text{PDEP}}$ ($N_e$) | kpoint | VBM (eV) | CBM (eV) | Gap (eV) |
|--------|------|--------|----------|----------|----------|
| ML BN | 4 | $9 \times 9$ | -7.6745 | -0.2216 | 7.453 |
| | 4 | $12 \times 12$ | -7.5602 | -0.2994 | 7.261 |
| | 4 | $15 \times 15$ | -7.4950 | -0.3471 | 7.148 |
| | 4 | $\infty$ (9/12) | -7.4132 | -0.3995 | 7.014 |
| | 4 | $\infty$ (12/15) | -7.3929 | -0.4089 | 6.984 |
| BL BN | 3 | $9 \times 9$ | -7.5200 | -0.2384 | 7.282 |
| | 4 | $9 \times 9$ | -7.5907 | -0.2944 | 7.296 |
| | 3 | $12 \times 12$ | -7.4172 | -0.2999 | 7.117 |
| | 4 | $12 \times 12$ | -7.4852 | -0.3558 | 7.129 |
| | 4 | $\infty$ (9/12) | -7.4327 | -0.4347 | 6.998 |
| TL BN | 3 | $9 \times 9$ | -7.4401 | -0.2829 | 7.157 |
| | 4 | $9 \times 9$ | -7.5081 | -0.3396 | 7.169 |
| | 3 | $12 \times 12$ | -7.3711 | -0.3122 | 7.059 |
| | 4 | $12 \times 12$ | -7.4404 | -0.3947 | 7.046 |
| | 4 | $\infty$ (9/12) | -7.3527 | -0.4316 | 6.921 |
| Bulk BN | 3 | $6 \times 6 \times 2$ | – – | | 6.009 |
| Graphane | 3 | $9 \times 9$ | -6.7315 | 0.0040 | 6.735 |
| | 4 | $9 \times 9$ | -6.8107 | -0.0674 | 6.743 |
| | 3 | $12 \times 12$ | -6.6449 | -0.0524 | 6.592 |
| | 3 | $\infty$ (9/12) | -6.5335 | -0.1249 | 6.409 |

SI Table VI: Computed $G_0W_0$ gaps and band edges of $h$-BN and graphane. The number of PDEP (eigenpotentials for the calculations of dielectric matrices) is 3 times of number of electrons ($3N_{\text{electrons}}$) and we used $4N_{\text{electrons}}$ to validate the convergence. To alleviate convergence issues of $G_0W_0$ with k-point sampling, we extrapolate our final results (highlighted in blue) from a 9x9 k-point mesh to a 12x12 k-point mesh (noted as "$\infty$(9/12)") based on the formula proposed in Ref. 8. This extrapolation gives a final gap accurate within 0.1 eV for monolayer $h$-BN system where we also tested a 15x15 k-point mesh to validate the convergence.



\end{document}